\documentclass{article}

\usepackage{amsmath,amsthm,amsopn,amsfonts,graphicx,amssymb,dsfont,color}
\usepackage{textcomp}
\usepackage{mathrsfs}
\usepackage{color}
\usepackage{geometry}
\usepackage{subfigure}
\usepackage{lineno}

\newcommand{\be}{\begin{equation}}
\newcommand{\ee}{\end{equation}}
\newcommand{\baa}{\begin{array}}
\newcommand{\eaa}{\end{array}}
\newcommand{\baco}{\left\{ \begin{array}}
\newcommand{\eaco}{\end{array} \right.}

\newcommand{\ds}{\displaystyle}

\def\L{\mathcal L}

\def\hI{\hat{\delta}}

\def\pos{posterior }
\def\pri{prior }

\title{Using early data to estimate the actual infection fatality ratio from COVID-19 in France}

\author{Lionel Roques$^{\hbox{\small{ a,*}}}$, Etienne Klein$^{\hbox{\small{ a}}}$, Julien Papaïx$^{\hbox{\small{ a}}}$,\\ Antoine Sar$^{\hbox{\small{ b}}}$ and Samuel Soubeyrand$^{\hbox{\small{ a}}}$ \\  \\\footnotesize{$^{\hbox{a }}$INRAE, BioSP, 84914, Avignon, France} \\\footnotesize{$^{\hbox{b }}$Medicentre Moutier, 2740 Moutier, Switzerland}\\ \footnotesize{$^{\hbox{* }}$Contact~: lionel.roques@inrae.fr} }

\date{}

\begin{document}

\maketitle

\begin{abstract}

\noindent\textit{Background.} The number of screening tests carried out in France and the methodology used to target the patients tested do not allow for a direct computation of the actual number of cases and the infection fatality ratio (IFR). The main objective was to estimate the actual number of people infected with COVID-19 during the observation window in France and to deduce the IFR.

\ 

\noindent\textit{Methods.} We develop a 'mechanistic-statistical' approach coupling a SIR epidemiological model describing the unobserved epidemiological dynamics, a probabilistic model describing the data acquisition process and a statistical inference method. 

\ 

\noindent\textit{Results.} The actual number of infected cases in France is probably higher than the observations: we find here a factor $\times 8$ ($95\%$-CI:~5-12) which leads to an IFR in France of $0.5\%$ (95\%-CI: $0.3-0.8$) based on hospital death counting data. Adjusting for the number of deaths in nursing homes, we obtain an IFR of $0.8\%$ (95\%-CI: $0.45-1.25$).

\ 

\noindent \emph{Conclusions.}  The IFR is consistent with previous findings in China (0.66\%) and lower than the value of 0.9$\%$ previously obtained in the UK.

\end{abstract}


\section{Background} 

The COVID-19 epidemic started in December 2019 in Hubei province, China. Since then, the disease has spread around the world reaching the pandemic stage, according to the WHO \cite{WHO20}, on March 11. The first cases were detected in France on January 24. The infection fatality ratio (IFR), defined as the number of deaths divided by the number of infected cases, is an important quantity that informs us on the expected number of casualties at the end of an epidemic, when a given proportion of the population has been infected. Although the data on the number of deaths from COVID-19 are probably accurate, the actual number of infected people in the population is not known. Thus, due to the relatively low number of screening tests that have been carried out in France (about 5 over $10,000$ people in France to be compared with 50 over $10,000$ in South Korea up to March 15, 2020; Sources: Santé Publique France and Korean Center for Disease Control) the direct computation of the IFR is not possible. Based on the PCR-confirmed cases in international residents repatriated from China on January 2020, \cite{VerOke20} obtained an estimate of the infection fatality ratio (IFR) of 0.66\% in China, and, adjusting for non-uniform attack rates by age, an IFR of 0.9$\%$ was obtained in the UK \cite{FerLay20}.

Using the early data (up to March 17) available in France, our objectives are: (1) to compute the IFR in France, (2) to estimate the number of people infected with COVID-19 in France, (3) to compute a basic reproduction rate $R_0$.

\section{Methods}

\paragraph{Data.} We obtained the number of positive cases and deaths in France, day by day from Johns Hopkins University Center for Systems Science and Engineering  \cite{DonDu20}.
The data on the number of tests carried out was obtained from Santé Publique France \cite{SPF20}. As some data (positive cases, deaths, number of tests) are not fully reliable (example: 0 new cases detected in France on March 12, 2020), we smoothed the data with a moving average over 5 days. Official data on the number of deaths by COVID-19 in France only take into account hospitalized people. About $728,000$ people in France live in nursing homes (EPHAD, source: DREES \cite{DREES20}). Recent data in the Grand Est region (source: Agence Régionale de Santé Grand Est \cite{ARS20}), report a total of 570 deaths in these nursing homes, which have to be added to the official count (1015 deaths on April 1st). 

\

\paragraph{Mechanistic-statistical model.}  This formalism, which is becoming standard in ecology \cite{RoqSouRou11, RoqBon16, AbbBon19} allows the analyst to couple a mechanistic model, here an ordinary differential equation model (ODE) of the SIR type, and uncertain, non-exhaustive data. To bridge the gap between the mechanistic model and the data, the approach uses a probabilistic model describing the data collection process. A statistical method is then used for the estimation of the parameters of the mechanistic model.

\

\noindent \textit{Mechanistic model.} The dynamics of the epidemics are described by the following SIR compartmental model \cite{Mur02}:
 \begin{equation}
\baco{l} \label{eq:SIR1}
 \ds S'(t)=- \frac{\alpha}{N} \, S(t) \, I(t), \vspace{1mm}\\
 \ds I'(t)=\frac{\alpha}{N} \, S(t) \, I(t) - \beta \, I(t), \vspace{1mm}\\
 \ds R'(t)= \beta \,  I(t),
\eaco
\end{equation}
with $S$ the susceptible population, $I$ the infected population, $R$ the recovered population (immune individuals) and $N=S+I+R$ the total population, supposed to be constant. The parameter $\alpha$ is the infection rate (to be estimated) and $1/\beta$ is the mean time until an infected becomes recovered. Based on the results in \cite{ZhoYu20}, the median period of viral shedding is 20 days, but the infectiousness tends to decay before the end of this period: the results in \cite{HeLau20} show that infectiousness starts from 2.5 days before symptom onset and declines within 7 days of illness onset. Based on these observations we assume here that $1/\beta=10$ days.

The initial conditions are $S(t_0)=N-1$, $I(t_0)=1$ and $R(t_0)=0$, where $N=67\, 10^6$ corresponds to the population size. The SIR model is started at some time $t=t_0$, which will be estimated and should approach the date of introduction of the virus in France (this point is shortly discussed at the end of this paper). The ODE system \eqref{eq:SIR1} is solved thanks to a standard numerical algorithm, using Matlab\textsuperscript{\tiny\textregistered} \textit{ode45} solver.

Next we denote by $D(t)$ the number of deaths due to the epidemic. Note that the impact of the compartment $D(t)$ on the dynamics of the SIR system and on the total population is neglected here. The dynamics of $D(t)$ depends on $I(t)$ trough the differential equation:
\begin{equation}\label{eq:D(t)}
    D'(t)=\gamma(t) \, I(t),
\end{equation}
with $\gamma(t)$ the mortality rate of the infecteds. 

\ 

\noindent \textit{Observation model.} We suppose that the number of cases tested positive on day $t$, denoted by $\hI_t$, follow independent binomial laws, conditionally on the number of tests $n_t$ carried out on day $t$, and on $p_t$ the probability of being tested positive in this sample:
\begin{equation} \label{eq:model_bino}
  \hI_t \sim Bi (n_t, p_t),
\end{equation}
The tested population consists of a fraction of the infecteds and a fraction of the susceptibles: $ n_t = \tau_1 (t) \, I(t) + \tau_2 (t) \, S (t) $. Thus, $$ p_t = \frac {\sigma\, \tau_1 (t) \, I (t)} {\tau_1 (t) \, I (t) + \tau_2 (t) \, S (t)} = \frac {\sigma\, I(t)} {I(t) + \kappa_t \, S (t)},$$ with $ \kappa_t: = \tau_2 (t) / \tau_1 (t) $, the relative probability of undergoing a screening test for an individual of type $ S $ {\it vs} an individual of type $ I $ (probability of being tested conditionally on being $ S $ / probability of being tested conditionally on being $ I $). We assume that the ratio $ \kappa $ does not depend on $ t $ at the beginning  of the epidemic (i.e., over the period that we use to estimate the parameters of the model).  The coefficient $\sigma$ corresponds to the sensitivity of the test. In most cases, RT-PCR tests have been used and existing data indicate that the sensitivity of this test using pharyngeal and nasal swabs is about $63-72\%$ \cite{WanXu20}. We take here $\sigma=0.7$ ($70\%$ sensitivity).

\paragraph{Statistical inference.} The data $\hI_t$ used to compute the MLE and the \pos distribution are those corresponding to the period from February 29 to March 17. The unknown parameters are  $\alpha$, $t_0$ and $\kappa$. The parameter $\gamma(t)$ is computed indirectly, using the estimated value of $I(t)$, the data on the mortality rate (assumed to be exact) and the relationship~\eqref{eq:D(t)}.

The likelihood $\mathcal{L}$, is defined as the probability of the observations (here, the increments $\{\hI_t\}$) conditionally on the parameters. Using the observation model \eqref{eq:model_bino}, and assuming that the increments $\hI_t$ are independent conditionally on the underlying SIR process and that the number of tests $n_t$ is known, we get:
$$\mathcal{L}(\alpha,t_0,\kappa):=P(\{\hI_t\} |\alpha,t_0,\kappa)=\prod_{t=t_i}^{t_f} \frac{n_t !}{(\hI_t)!(n_t-\hI_t)!} p_t^{\hI_{t}}\, (1-p_t)^{n_t-\hI_t},$$with $t_i$ the date of the first observation and $t_f$ the date of the last observation. In this expression $\mathcal{L}(\alpha,t_0,\kappa)$ depends on $\alpha,t_0,\kappa$ through $p_t$.

The maximum likelihood estimator (MLE, i.e., the parameters that maximize $\L $), is computed using the BFGS constrained minimization algorithm, applied to $ - \ln (\L) $, via the Matlab \textsuperscript{\tiny \textregistered} function \textit{fmincon}. In order to find a global maximum of $\L $, we apply this method starting from random initial values for $\alpha,t_0, \kappa$ drawn uniformly in the following intervals:
 \begin{equation}
\baco{l} \label{eq:prior}
 \alpha \in (0,1),\\t_0 \in (1,50), \hbox{ (January 1st - February 19th)}\\\kappa \in (0,1).
\eaco
\end{equation}
The minimization algorithm is applied to 10000 random initial values of the parameters.

To assess model fit, we compare the observations with expectation of the observation model associated with the MLE, $n_t \, p_t^*$ (expectation of a binomial) with $$p_t^* = \frac {\sigma \, I^* (t)} {I^* (t) + \kappa^ * \, S^* (t)},$$and $I^* (t) $, $ S^* (t) $ the solutions of the system \eqref{eq:SIR1} associated with the MLE.

The \pos distribution of the parameters $ (\alpha, t_0, \kappa) $ is computed with a Bayesian method, using uniform \pri distributions in the intervals given by~\eqref{eq:prior}. This \pos distribution corresponds to the distribution of the parameters conditionally on the observations:
$$P(\alpha, t_0, \kappa | \{\hI_t\}) = \frac {\L(\alpha, t_0, \kappa) \, \pi (\alpha,t_0,\kappa)} {C},$$where $\pi(\alpha, t_0, \kappa) $ corresponds to the \pri distribution of the parameters (therefore uniform) and $C$ is a normalization constant independent of the parameters. The numerical computation of the \pos distribution  (which is only carried out for French data) is performed with a Metropolis-Hastings (MCMC) algorithm, using 4 independent chains, each of which with $ 10^6 $ iterations, starting from random values close to the MLE.

\paragraph{Computation of the infection fatality ratio and of $R_0$.} The IFR corresponds to the fraction of the infected who die, that is $\gamma (t) / (\gamma(t) + \beta) $. Given the (estimated) population $I$, the term $ \gamma (t) $ is computed using the formula  \eqref{eq:D(t)} and the mortality data. With SIR systems of the form~\eqref{eq:SIR1}, the basic reproduction rate $R_0$ can be computed directly, based on the formula $R_0=\alpha/\beta$ \cite{Mur02}. When $R_0<1$, the epidemic cannot spread in the population. When $R_0>1$, the infected compartment $I$ increases as long as $R_0 \, S>N=S+I+R$.

\section{Results}

\begin{figure}
\center
\includegraphics[trim={0 2.5cm 0 4cm},clip,width=0.8\textwidth]{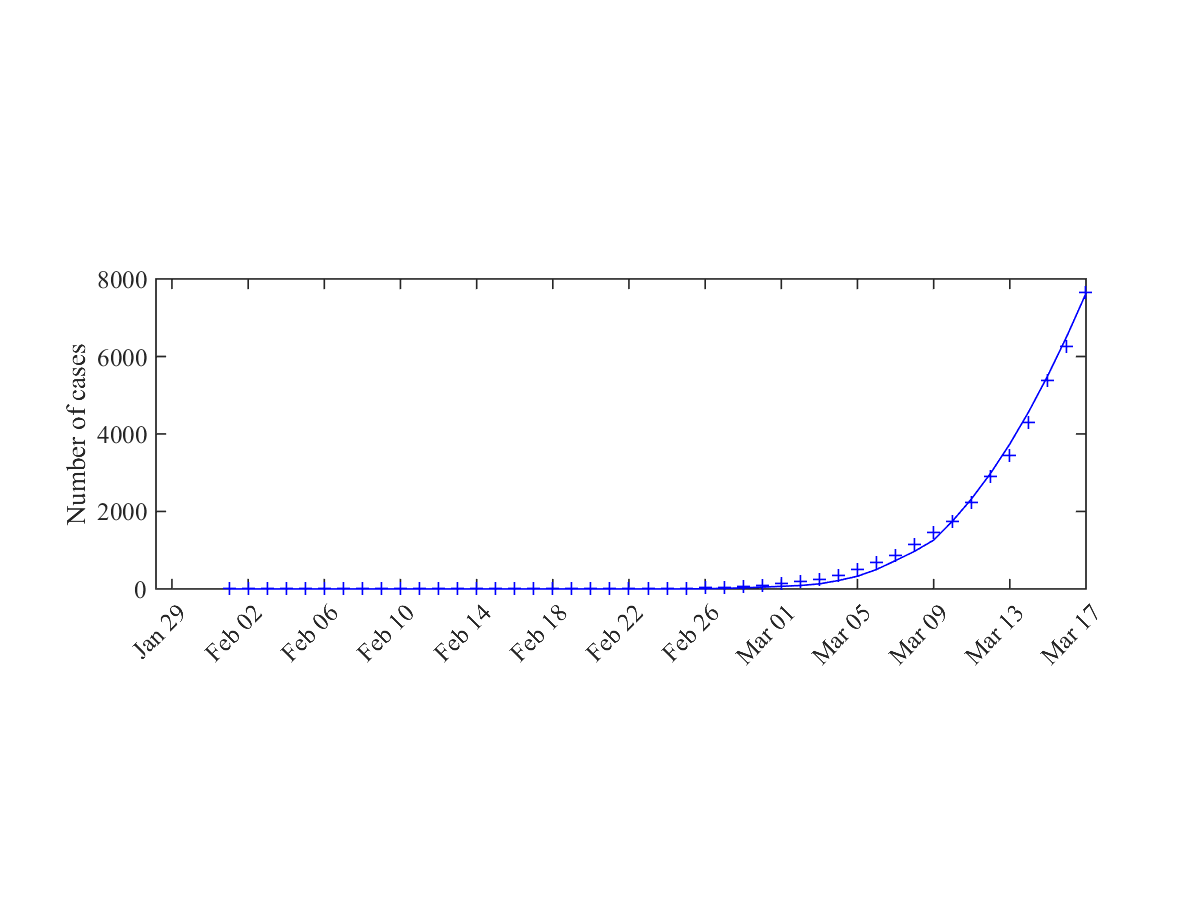}
\caption{{\bf Expected number of observed cases associated with the MLE {\it vs} number of cases actually detected (total cases).} The curves corresponds to the expected observation $ n_t \, p_t ^ * $ given by the model, and the crosses correspond to the data (cumulated values of $\hI_t$). }
 \label{fig:fit}
\end{figure}

\noindent \textit{Model fit.} Fig.~\ref{fig:fit} compares the expectation of the observation model associated with the MLE with the actual observations. We get a good match between this expectation $n_t \, p_t^*$ and the data.

\

\begin{figure}
\center
\includegraphics[trim={0 3cm 0 4cm},clip,width=0.8\textwidth]{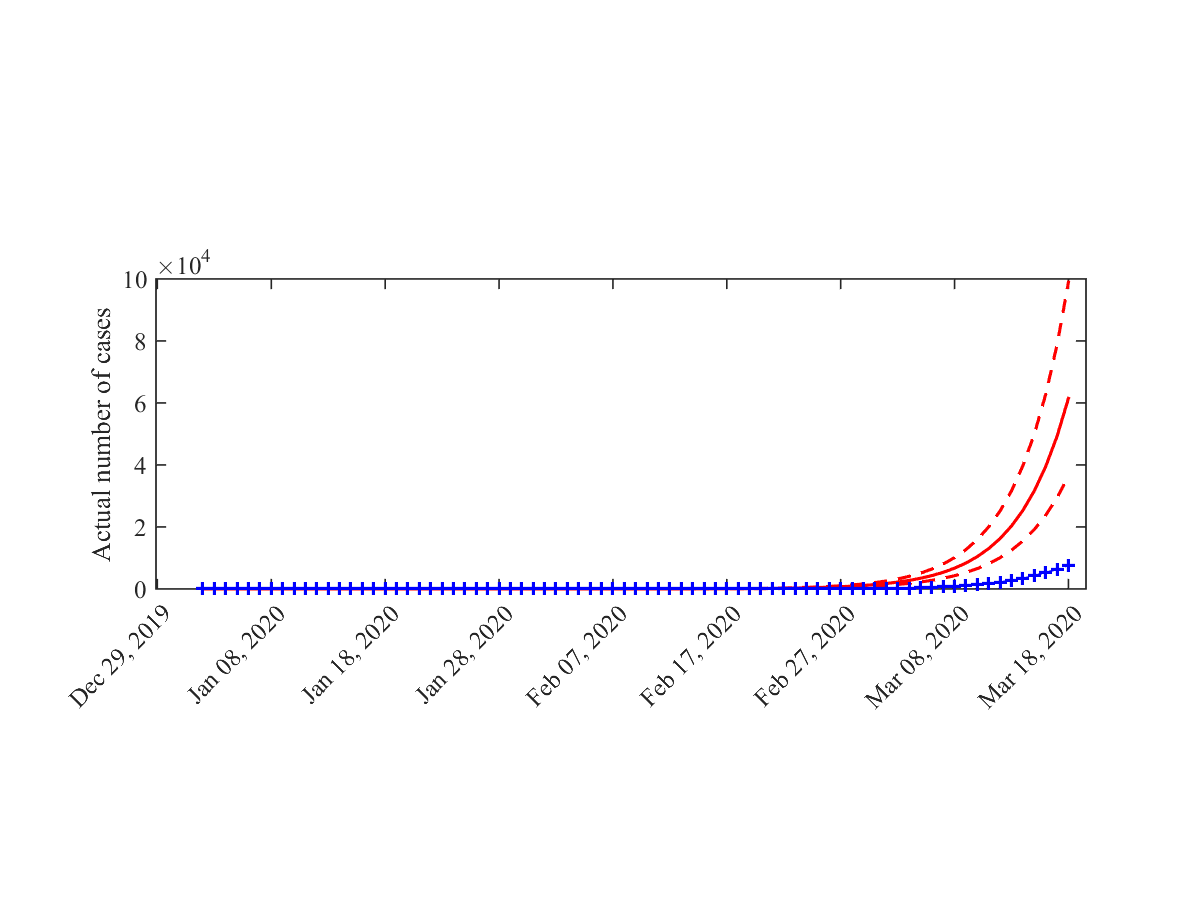}
%\subfigure[$\delta(t)=\delta_{max} \sin^2(\omega\, t)$]{\includegraphics[width=0.49\textwidth]{{fig_sin2_U=10Uc_dmax=10sqrtlambda_N=1000}.png}}
\caption{{\bf Distribution of the cumulated number of infected cases ($I(t)+R(t)$) across time.} Solid line: average value obtained from the \pos distribution of the parameters. Dotted curves: 0.025 and 0.975 pointwise \pos quantiles. Blue crosses: data (cumulated values of $\hI_t$).}
 \label{fig:IR}
\end{figure}

\noindent \textit{Infection fatality rate.} Using the \pos distribution of the model parameters (the pairwise distributions are available as Supplementary Material, see Fig.~S1), we can compute the daily distribution of the actual number of infected peoples. Using this information we thus obtain, on March 17, an IFR of $0.5\%$ (95\%-CI: $0.3-0.8$). The estimated distribution of IFR is relatively stable over time, see Fig.~S2 in Supplementary Material.  Additionally, the distribution of the cumulated number of infected cases ($I(t)+R(t)$) across time is presented in Fig.~\ref{fig:IR}. We observe that it is much higer than the total number of observed cases (compare with Fig.~\ref{fig:fit}). The average estimated ratio between the actual number of individuals that have been infected and observed cases ($ (I(t)+R(t)) / \Sigma \hI_t $, with $\Sigma\hI_t$ the sum of the observed infected cases at time $ t $) is $8$ ($95\%$-CI:~5-12) over the considered period. 

\

\noindent \textit{Taking into account the data in the nursing homes.} The above computation of the IFR is based on the official counting of deaths by COVID-19 in France, which does not take into account the number of deaths in nursing homes. Based on the local data in Grand Est region, we infer that the IFR that we computed has been underestimated by a factor about $(1015+570)/1015\approx 1.6$, leading to an adjusted IFR of $0.8\%$ (95\%-CI: $0.45-1.25$).

\

\noindent \textit{Basic reproduction rate.} We computed the  marginal \pos distribution of the basic reproduction rate $R_0$.  This leads to a mean value of $R_0$ of 3.2 ($95\%$-CI:~3.1-3.3). The full distribution is available as Supplementary Material (Fig.~S3).

\section{Discussion.}

\noindent \textit{On the IFR and the number of infecteds.} The actual number of infected individuals in France is probably much higher than the observations (we find here a factor $\times 8$), which leads at a lower mortality rate than that calculated on the basis of the observed cases: we find here an IFR of $0.5\%$ based on hospital death counting data, to be compared with a case fatality rate (CFR, number of deaths over number of diagnosed cases) of $2\%$. Adjusting for the number of deaths in the nursing homes, we obtained an IFR of $0.8\%$.
 These values for the IFR are consistent with the findings of \cite{VerOke20}  (0.66\% in China), and slightly lower than the value  of 0.9$\%$ previously obtained in the UK \cite{FerLay20}. If the virus led to contaminate $80\% $ of the French population \cite{FerLay20}, the total number of deaths to deplore in the absence of variation in the mortality rate (increase induced for example by the saturation of hospital structures, or decrease linked to better patient care) would be $336 , 000$  (95\%-CI ~: $192,000-537,000$), excluding the number of deaths in the nursing homes. This estimate could be corroborated or invalidated when $80\%$ of the population will be infected, eventually over several years, assuming that an infected individual is definitively immunized. It has to be noted that measures of confinement or social distancing can decrease both the percentage of infected individuals in the population and the degree of saturation of hospital structures.

%\

%\noindent \textit{On the differences between France and South Korea.} The mechanistic-statistical SIR model achieves a satisfactory goodness-of-fit for French data, but does not capture the decline in the number of cases observed in South Korea. The difference between the dynamics predicted by the SIR model and the South Korean data is probably linked to a different management of the epidemic in Korea, having a strong impact on the epidemic dynamics (more important screening, tracing, social distancing in South Korea).

\

\noindent \textit{On the value of $R_0$.} The estimated distribution in France is  high compared to recent estimates (2.0-2.6, see \cite{FerLay20}) but consistent with the findings in \cite{ZhaLin20} (2.24-3.58). A direct estimate, by a non-mechanistic method, of the parameters $ (\rho, t_0) $ of a model of the form $ \hI_t = e ^ {\rho \, (t-t_0)} $ gives $ t_0 = $ 36 (February 5) and $ \rho = $ 0.22. With the SIR model, $ I '(t) \approx I \, (\alpha - \beta) $ for small times ($ S \approx N $), which leads to a growth rate equal to $ \rho \approx \alpha- \beta $, and a value of $ \alpha \approx 0.32 $, that is to say $ R_0 = 3.2$, which is consistent with our distribution of $R_0$. Note that we have assumed here a infectiousness period of $10$ days. A shorter period would lead to a lower value of $R_0 $.

\

\noindent \textit{On the uncertainty linked to the data.} The uncertainty on the actual number of infected and therefore the mortality rate are very high. We must therefore interpret with caution the inferences that can be made based on the data we currently have in France. In addition, we do not draw forecasts here: the future dynamics will be strongly influenced by the containment measures that will be taken and should be modeled accordingly.

\

\noindent \textit{On the hypotheses underlying the model.} The data used here contain a limited amount of information, especially since the observation period considered is short and corresponds to the initial phase of the epidemic dynamics, which can be strongly influenced by discrete events. This limit led us to use a particularly parsimonious model in order to avoid problems of identifiability for the parameters. The assumptions underlying the model are therefore relatively simple and the results must be interpreted with regard to these assumptions. For instance, the date of the introduction $ t_0 $ must be seen as an {\it efficient} date of introduction for a dynamics where a single introduction would be decisive for the outbreak and the other (anterior and posterior) introductions would have an insignificant effect on the dynamics.

\section*{Acknowledgements}
This work was funded by INRAE.

\section*{Author contributions statement}

L.R., E.K.K., J.P., A.S. and S.S. conceived the model and designed the statistical analysis. L.R. and S.S. wrote the paper, L.R. carried out the numerical computations. All authors reviewed the manuscript.

\section*{Competing interests}

All authors report no conflict of interest relevant to this article.

\clearpage

\section*{Supplementary Material}

\begin{figure}[h]
\center
\subfigure{\includegraphics[width=0.48\textwidth]{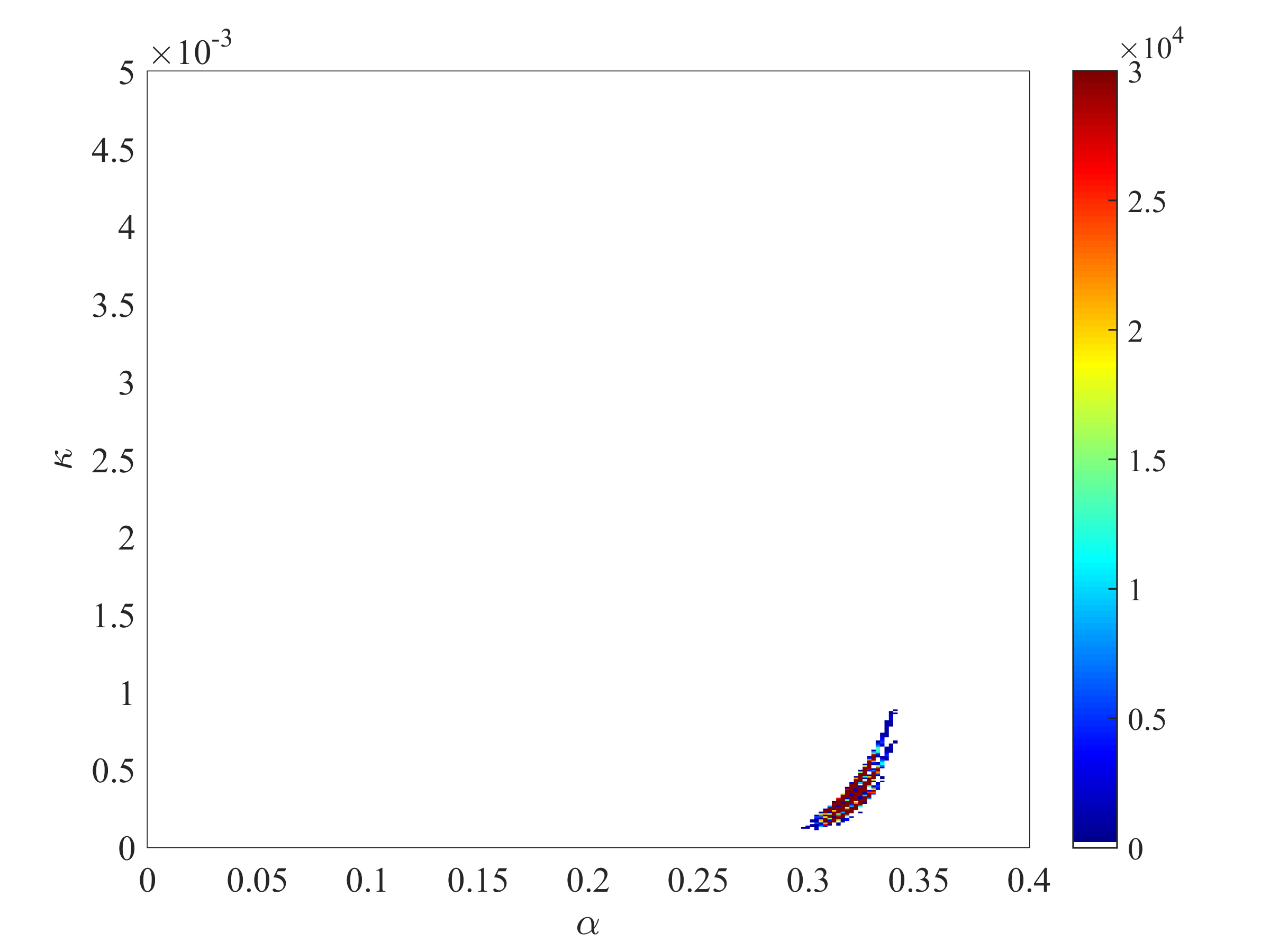}}
\subfigure{\includegraphics[width=0.48\textwidth]{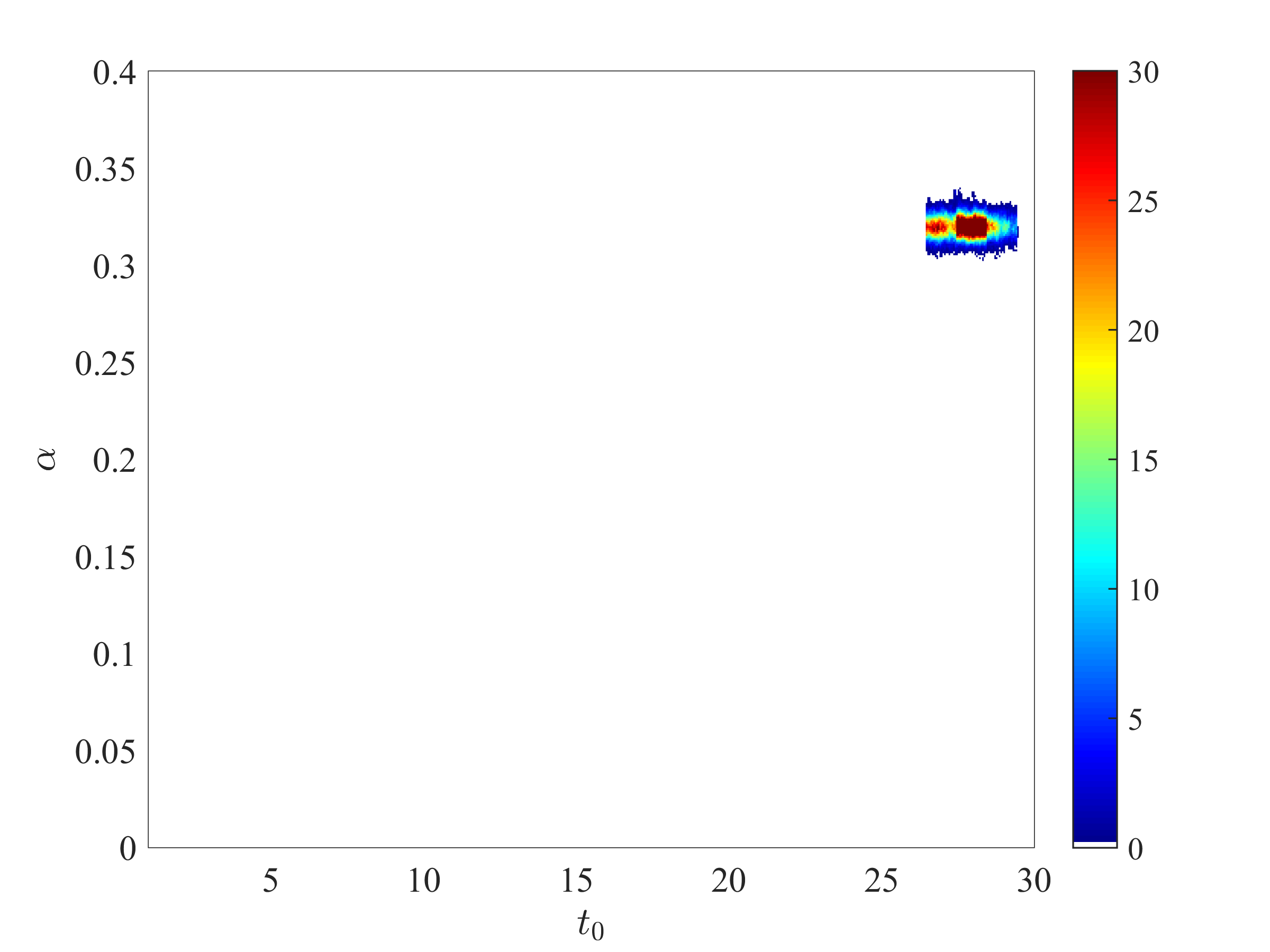}}
\subfigure{\includegraphics[width=0.48\textwidth]{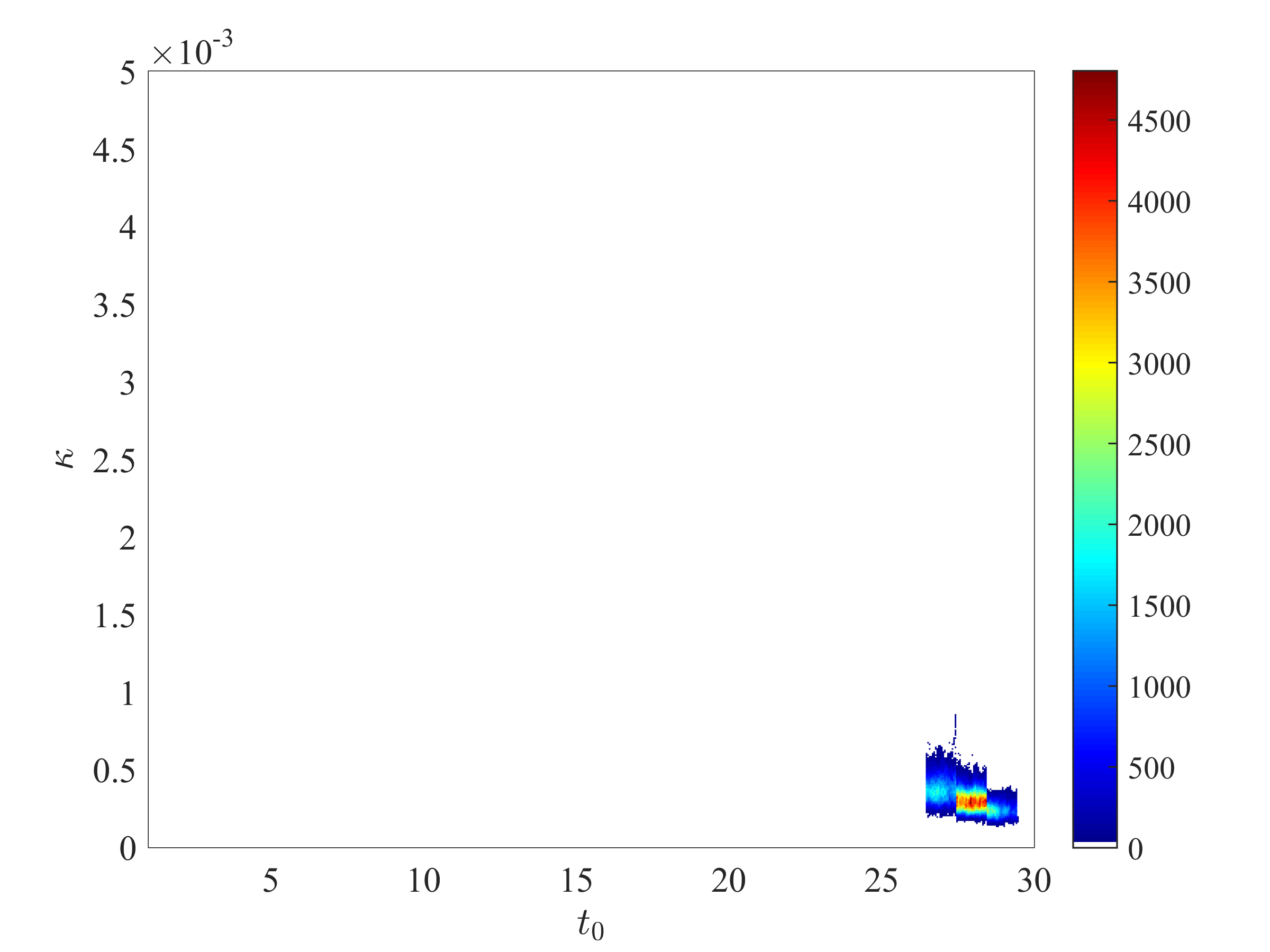}}
\caption{{\bf Joint \pos distributions of  $(\alpha,\kappa)$, $(t_0,\alpha)$ and $(t_0,\kappa)$.}}
 \label{fig:jointe}
\end{figure}

\begin{figure}
\center
\includegraphics[trim={0 3cm 0 4cm},clip,width=0.8\textwidth]{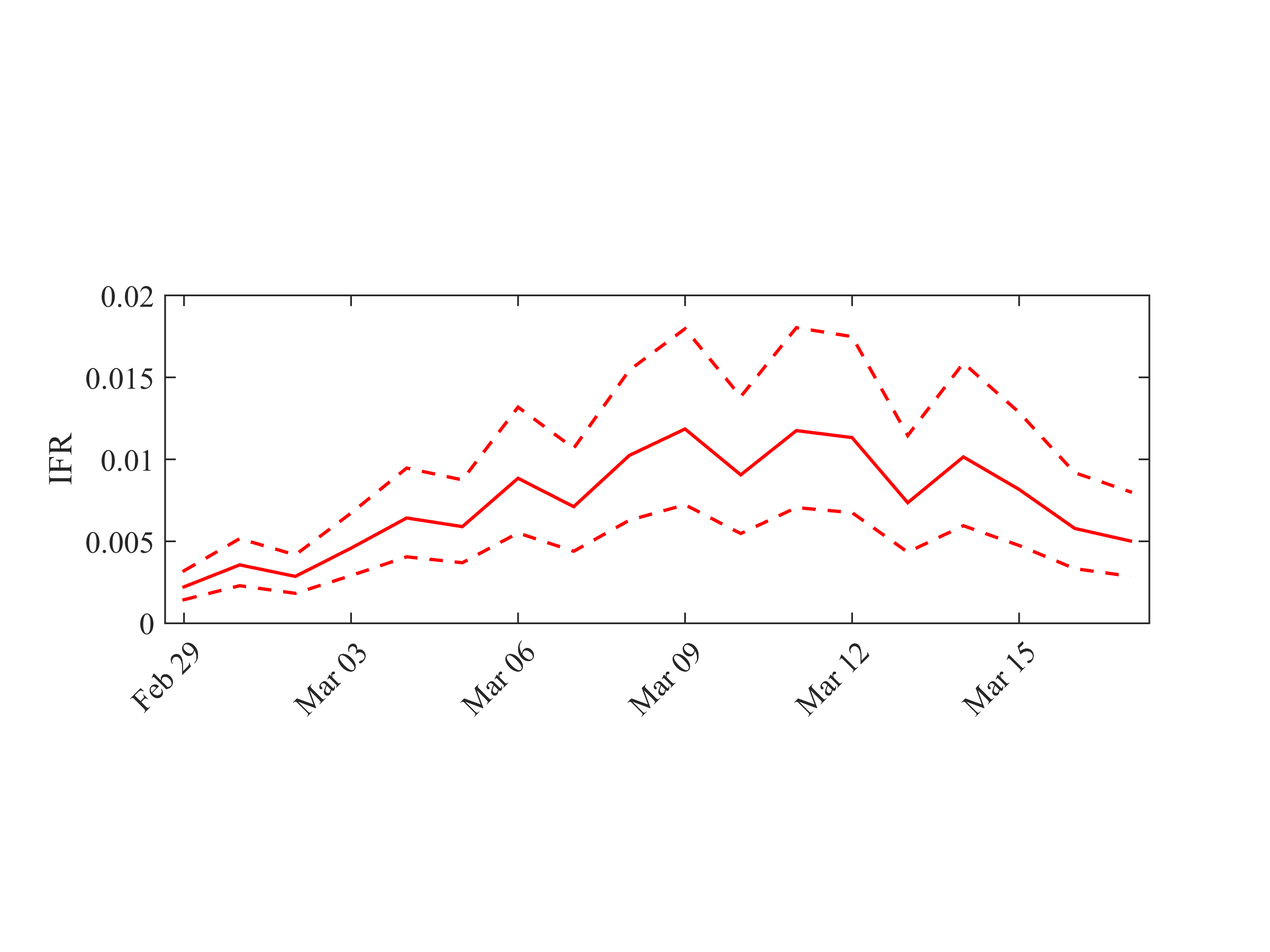}
%\subfigure[$\delta(t)=\delta_{max} \sin^2(\omega\, t)$]{\includegraphics[width=0.49\textwidth]{{fig_sin2_U=10Uc_dmax=10sqrtlambda_N=1000}.png}}
\caption{{\bf Dynamics of the IFR in France.} Solid line: average value obtained from the \pos distribution of the parameters. Dotted curves: 0.025 and 0.975 pointwise quantiles.}
 \label{fig:D}
\end{figure}

\begin{figure}
\center
\includegraphics[width=0.5\textwidth]{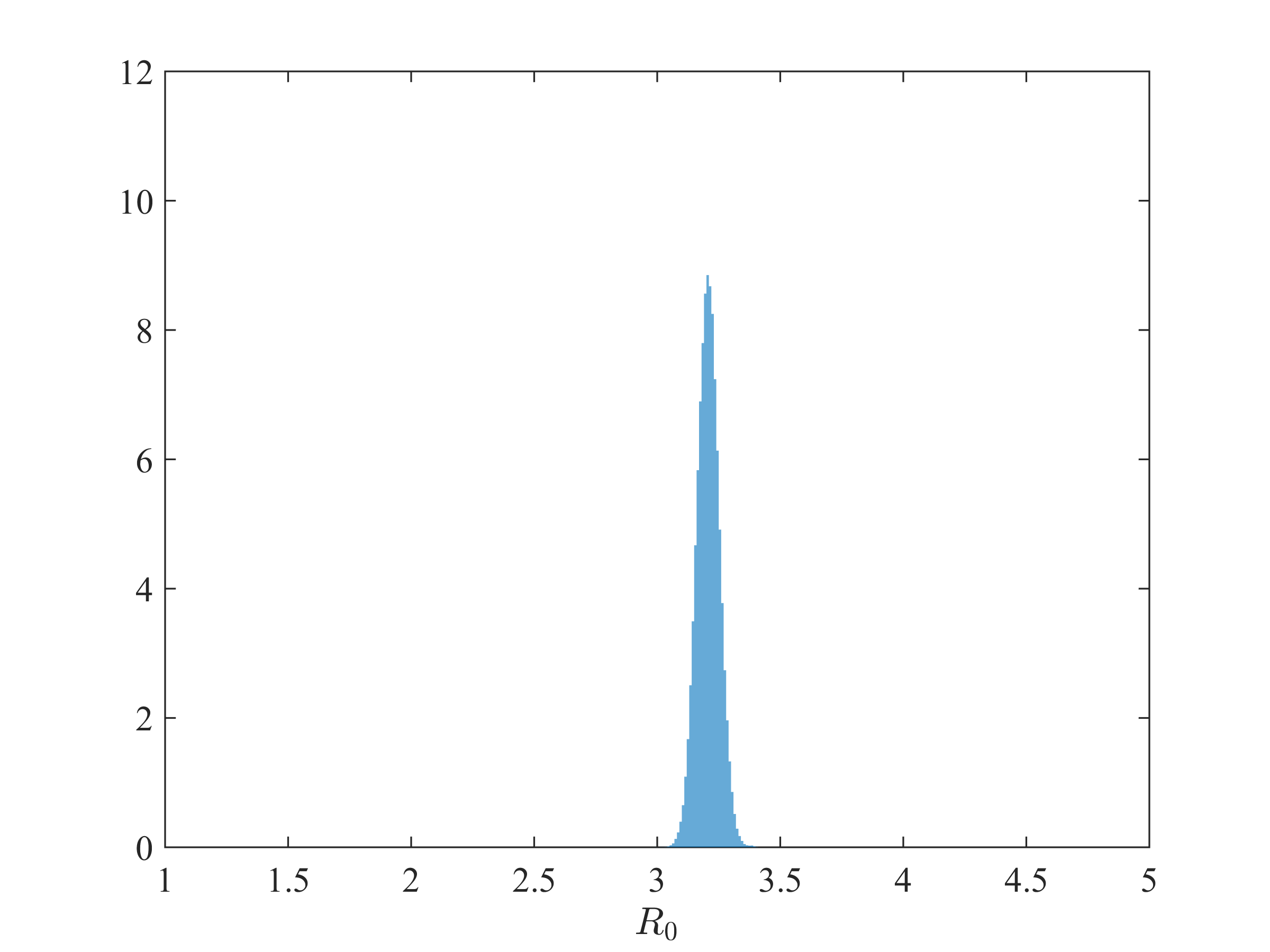}
%\subfigure[$\delta(t)=\delta_{max} \sin^2(\omega\, t)$]{\includegraphics[width=0.49\textwidth]{{fig_sin2_U=10Uc_dmax=10sqrtlambda_N=1000}.png}}
\caption{{\bf Posterior distribution of the basic reproduction rate $R_0$ in France.}}
 \label{fig:R0}
\end{figure}

\begin{itemize}
    \item[-] The joint \pos distributions of the three pairs of parameters $(\alpha,\kappa)$, $(t_0,\alpha) $ and $(t_0,\kappa)$
are depicted in Fig.~\ref{fig:jointe}.
    \item[-] The dynamics of the estimated distribution of the IFR are depicted in Fig.~\ref{fig:D}.
    \item[-] The marginal posterior distribution of $R_0$ is depicted in Fig.~\ref{fig:R0}.
\end{itemize}

\end{document}